\documentclass[12pt]{article}
\usepackage{epsf,amssymb}

\newcommand{\figone}
{
\begin{figure}[tbh]
\centerline{\epsfxsize=8cm \epsfbox{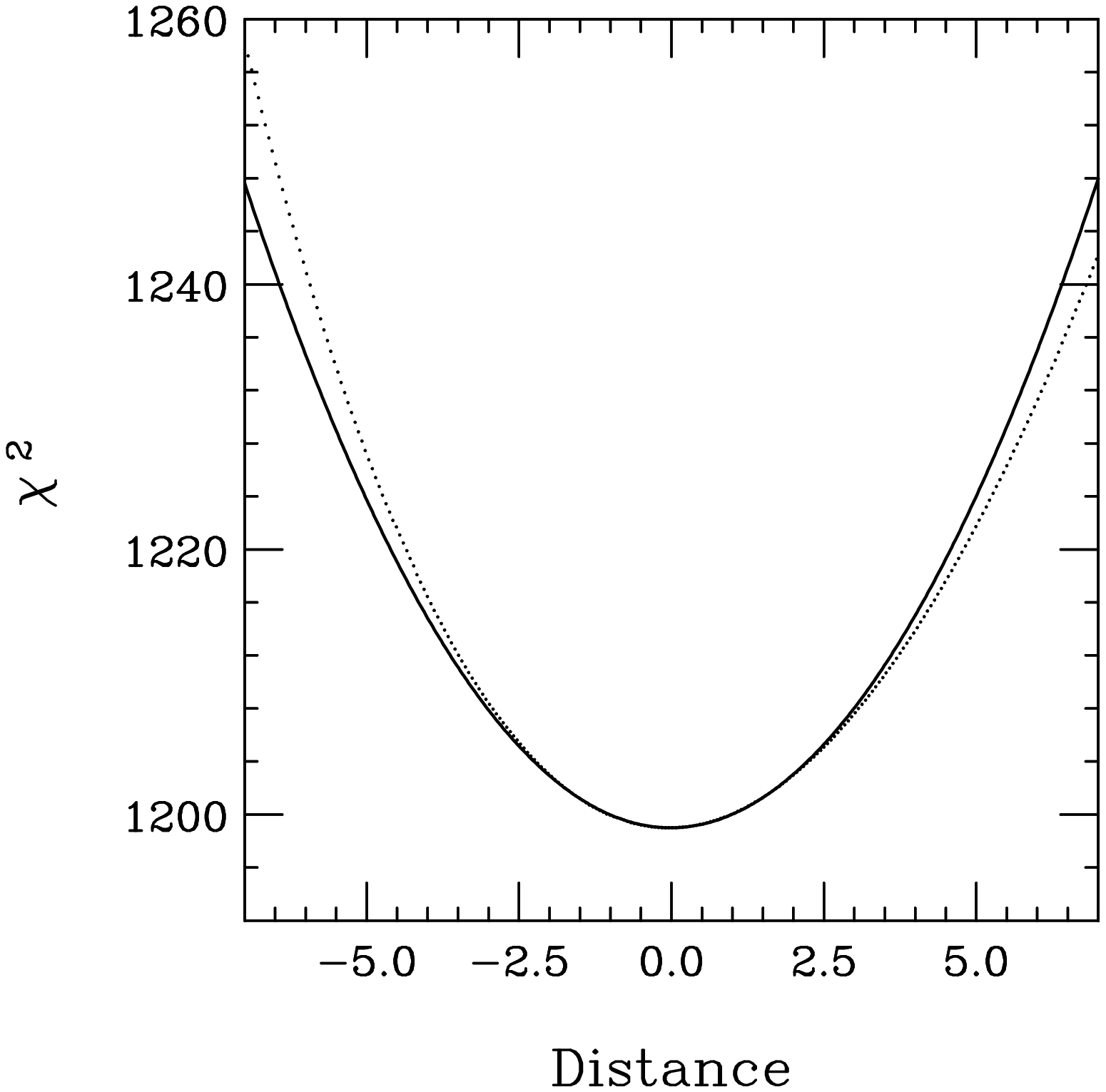} }
\caption{
Variation of $\chi^{2}$ with distance along a typical
direction in parameter space.
The dotted curve is the exact $\chi^{2}$ and the solid curve
is the quadratic approximation based on the Hessian.
The quadratic form is seen to be a rather good approximation
over the range shown.
}
\label{fig:one}
\end{figure}
}

\newcommand{\figtwo}
{
\begin{figure}[tbh]
\centerline{\epsfxsize=10cm \epsfbox{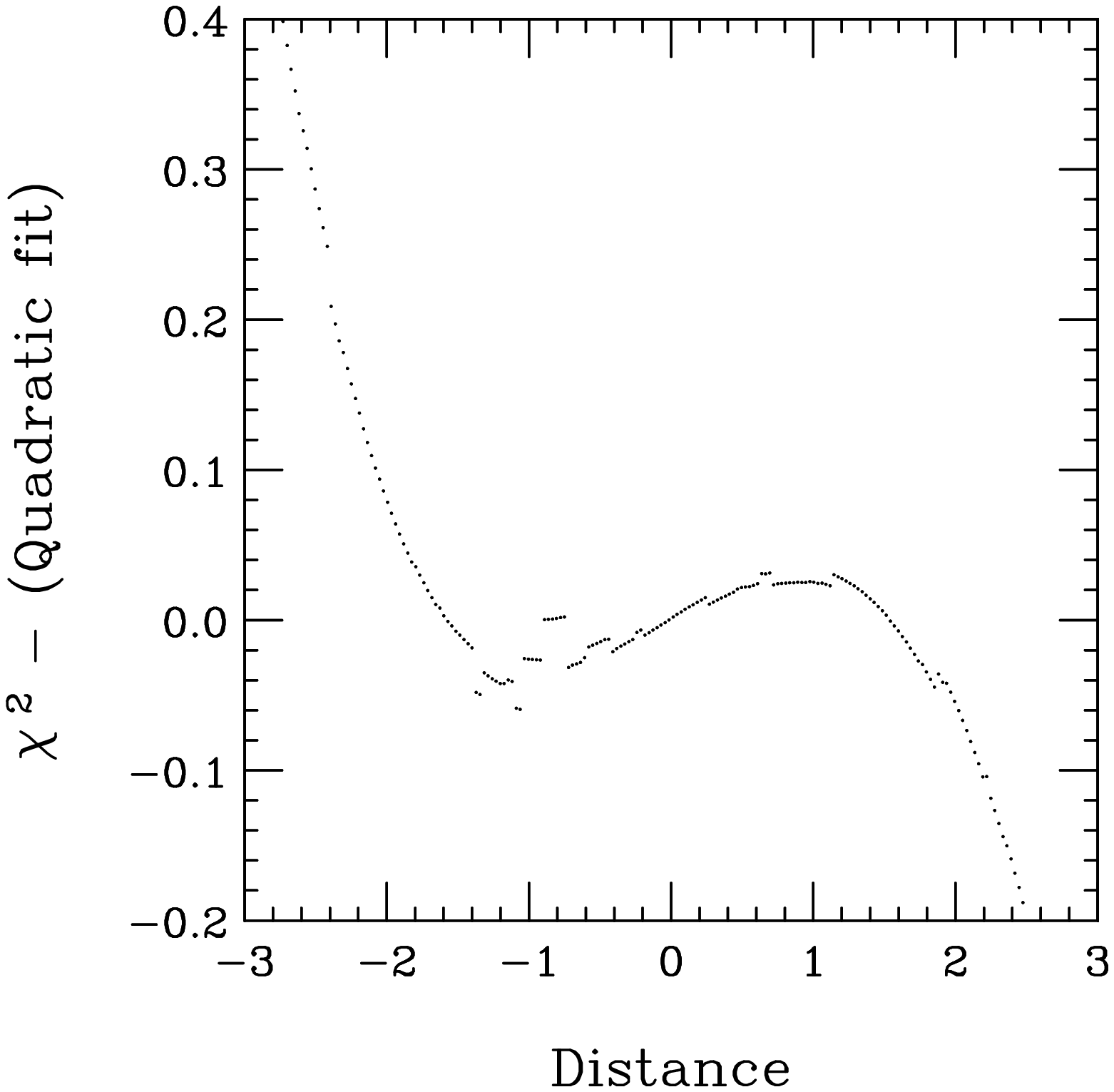} }
\caption{
Difference between $\chi^{2}$ and its quadratic approximation
(\ref{eq:taylor}), both of which are shown in Fig.\ \ref{fig:one}.
A cubic contribution can be seen, along with a
noticeable amount of numerical noise.  The fine structure
revealed here is small compared to the main variation of
$\chi^{2}$ itself, which rises by $20$ over the region
shown, as can be seen in Fig.~1.
}
\label{fig:two}
\end{figure}
}

\newcommand{\figthree}
{
\begin{figure}[tbh]
\centerline{\epsfxsize=8cm \epsfbox{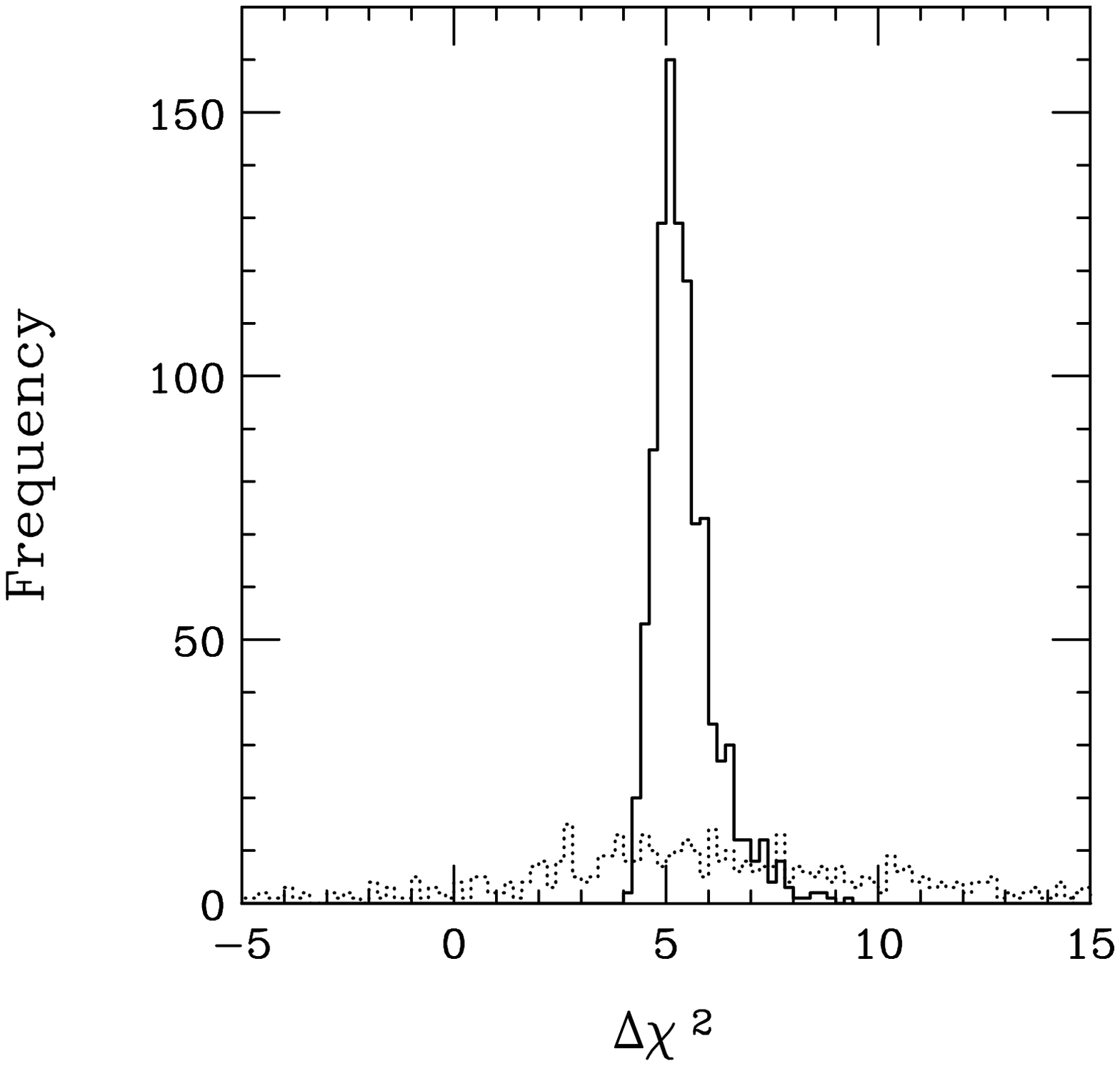} }
\caption{
Frequency distribution of $\Delta \chi^2$ according to the
Hessian approximation (\ref{eq:taylor}) for displacements
in random directions for which the true value is
$\Delta \chi^2 = 5.0\,$.  {\it Solid histogram}: using Hessian
calculated by iterative method of Section 3; {\it Dotted
histogram}: using Hessian calculated by {\footnotesize MINUIT}.
}
 \label{fig:three}
\end{figure}
}

\newcommand{\figfour}
{
\begin{figure}[tbh]
\centerline{\epsfxsize=8cm \epsfbox{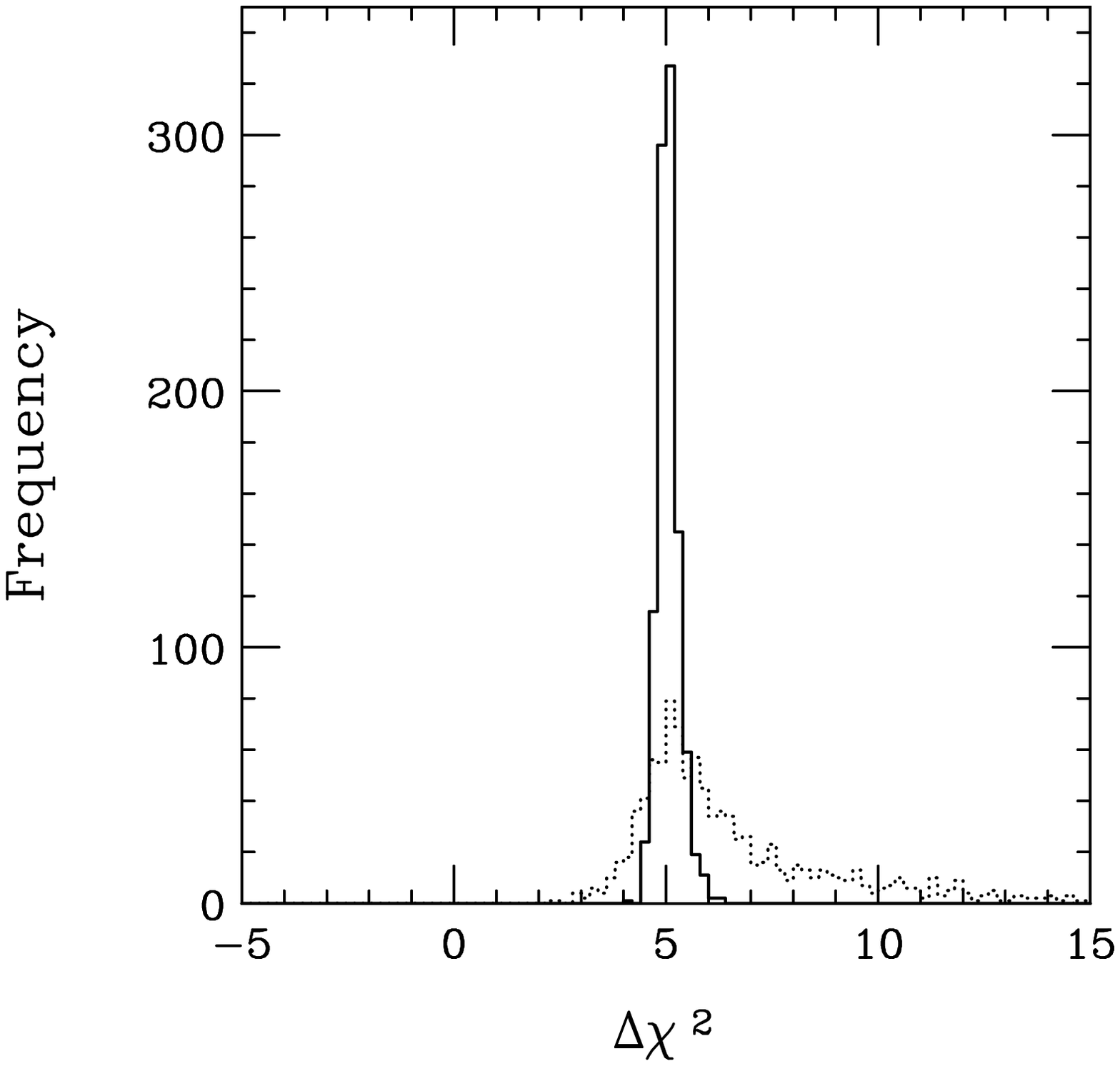} }
\caption{
Same as Fig.\ \ref{fig:three}, except that the displacements are
restricted to the parameter subspace spanned by the 10 steepest
directions.
}
 \label{fig:four}
\end{figure}
}

\newcommand{\figfive}
{
\begin{figure}[tbh]
\centerline{\epsfxsize=8cm \epsfbox{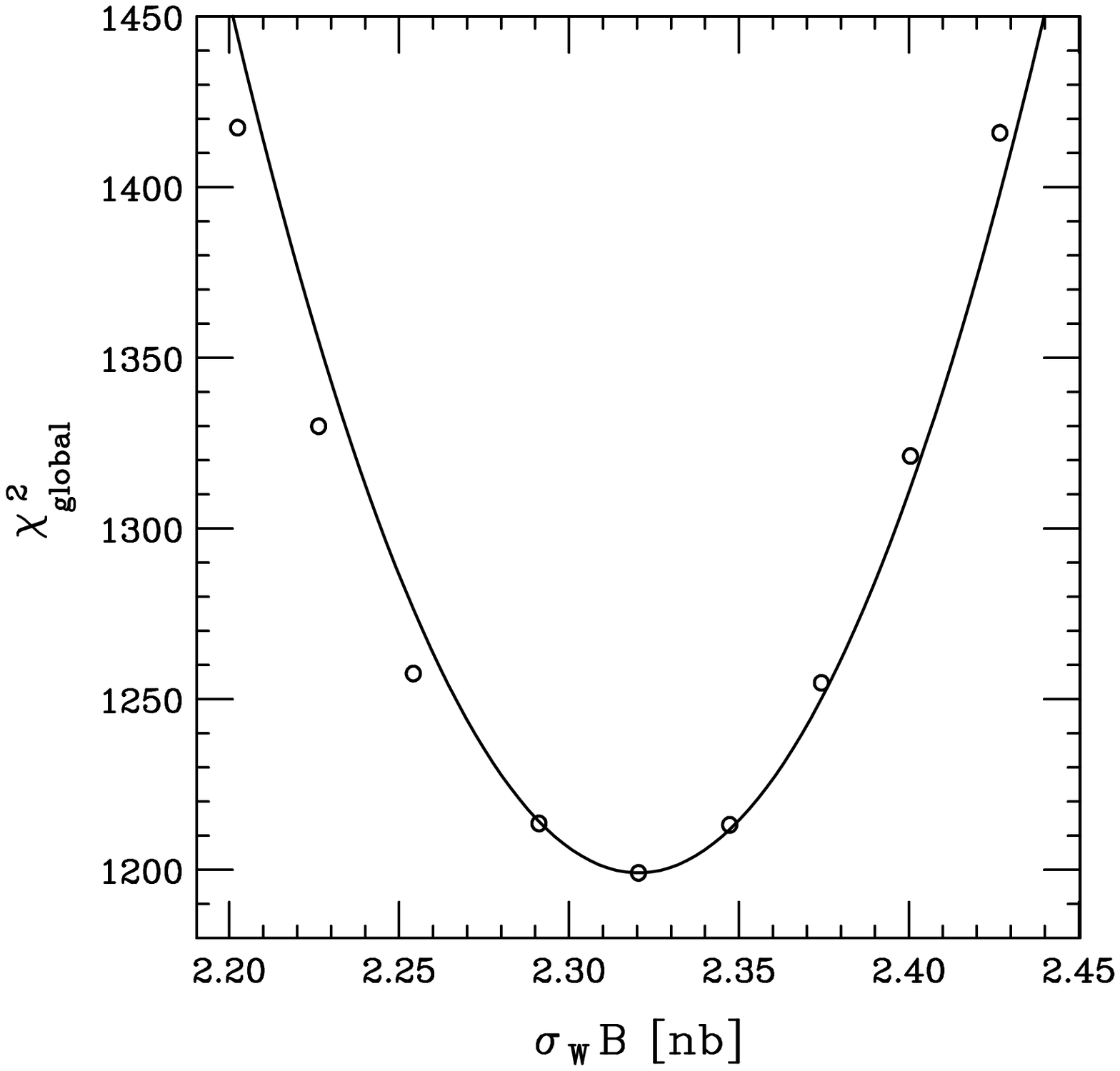} }
\caption{
Minimum $\chi^{2}$ as a function of the predicted cross
section for $W^{\pm}$ production in $p\overline{p}$ collisions.
{\it Parabolic curve} is the
prediction of the iteratively improved Hessian method.
{\it Points} are from the Lagrange multiplier method.
}
\label{fig:five}
\end{figure}
}

\newcommand{\DATE}  {\today}

\newcommand{\PPrtNo}
{
MSU-HEP-07100 \\
CERN-TH/2000-249
}
\newcommand{\TITLE}
{
Multivariate Fitting and the Error Matrix \\
in Global Analysis of Data
}

\newcommand{\AUTHORS}
{
J.\ Pumplin,$^a$ D.\ R.\ Stump,$^a$ and W.\ K.\ Tung$^{a,b}$
}
\newcommand{\INST}
{
$^a$ Department of Physics and Astronomy \\
         Michigan State University \\
         East Lansing, MI 48824 USA

\vspace{1cm}
$^b$ Theory Division \\
     CERN \\
     CH-1211 Geneva 23, Switzerland
}
\newcommand{\ABSTRACT}
{
When a large body of data from diverse experiments is analyzed
using a theoretical model with many parameters, the
standard error matrix method and the general tools for
evaluating errors may become inadequate.  We present an
iterative method that significantly improves the reliability
of the error matrix calculation.  To obtain even better
estimates of the uncertainties on predictions of physical
observables, we also present a Lagrange multiplier method that explores
the entire parameter space and
avoids the linear approximations assumed in conventional
error propagation calculations.
These methods are illustrated by an example
from the global analysis of parton distribution functions.
}

\newenvironment{Simlis}[2][$\bullet$]
{\begin{list}{#1}
 {
  \settowidth{\labelwidth}{#1}
  \setlength{\labelsep}{0.5em}
  \setlength{\leftmargin}{#2}
  \setlength{\rightmargin}{0em}
  \setlength{\itemsep}{0ex}
  \setlength{\topsep}{0ex}
 }
}
{\end{list}}
\begin{document}

\begin{titlepage}

\begin{tabular}{l}
\noindent\DATE
\end{tabular}
\hfill
\begin{tabular}{l}
\PPrtNo
\end{tabular}

\vspace{1cm}

\begin{center}
\renewcommand{\thefootnote}{\fnsymbol{footnote}}
{
\LARGE \TITLE
}

\vspace{1.25cm}
{\large  \AUTHORS}

\vspace{1.25cm}

\INST
\end{center}

\vfill

\ABSTRACT                 

\vfill

\newpage
\end{titlepage}



\section{Introduction}

\label{sec:Introduction}

The subject of this paper is a problem that arises when a large
body of data from diverse experiments is analyzed according to a
theoretical model that has many adjustable parameters.
Consider a generic data fitting problem based on experimental
measurements $\{D_{I},\,I=1,\dots,N\}$ with errors $\{\sigma_{I}\}$.
The data are to be compared to predictions
$\{T_{I}\}$ from a theoretical model with unknown
parameters $\{a_{i}, \, i=1,\dots,n\}$.
A common technique for comparing data with theory is to compute
the $\chi^{2}$ function defined by
\begin{equation}
\chi^{2}=\sum_{I=1}^{N}
\left(\frac{D_I - T_I}{\sigma_I}\right)^{2} \, ,
\label{eq:ChisqDef}
\end{equation}
or a generalization of that formula if
correlations between the errors are known in terms of a set of
correlation matrices.  The physics objectives are (i) to find the
best estimate of the parameters $\{a_{i}\}$ and their uncertainties,
and (ii) to predict the values and uncertainties of physical
quantities $\{X^{(\alpha)}, \, \alpha =1,2,\dots\}$ that are functions
of the $\{a_{i}\}$.

If the errors are randomly distributed, and the correlations well
determined, then standard statistical methods of $\chi^{2}$
minimization \cite{textbook,CSCTEQ} apply, and established
fitting tools like the CERN Library program
{\footnotesize MINUIT} \cite{minuit}
can be employed.  However, real problems are often more complex.
This is particularly true in a ``global analysis,'' where the large
number of data points $\{D_{I}\}$ do not come from a uniform set of
measurements, but instead consist of a collection of results from
many experiments on a variety of physical processes, with diverse
characteristics and errors.  The difficulties are compounded if
there are unquantified theoretical uncertainties, if the number of
theoretical parameters $n$ is large, or if the best parametrization
cannot be uniquely defined \textit{a priori}.  All of these
difficulties arise in the global analysis of hadronic
parton distribution functions (PDFs) \cite{CTEQn,MRST,cteq5},
which originally motivated this investigation.
Several groups have addressed the question of estimating errors
for the PDF determinations \cite{runII,Botje,Barone,Giele}.
But the problem is clearly more general than that application.

Of the many issues that confront a global analysis,
we address in this paper two, for which we have been able to
significantly improve on the traditional treatment. The
improvements allow a
more reliable determination of the uncertainties of $\{a_{i}\}$
and $\{X^{(\alpha)}\}$ in complex systems for which conventional methods
may fail. To define these problems, we assume
the system can be described by a global fitting function
$\chi_{\rm global}^{2}$, or $\chi^{2}$ for short, that characterizes
the goodness-of-fit for a given set of theory parameters $\{a_{i}\}$.
This $\chi^{2}$ distills all available information on the theory and
on the global experimental data sets, including their errors and
correlations.  One finds the minimum value $\chi_{0}^{2}$ of $\chi^{2}$,
and the best estimate of the theory parameters
are the values $\left\{ a_{i}^{0}\right\}$ that produce that minimum.
The dependence of  $\chi^{2}$ on $\{a_{i}\}$ near the minimum
provides information on the uncertainties in the $\{a_{i}\}$.
These are usually characterized by the error matrix and its inverse,
the Hessian matrix $H_{ij}$, where one assumes that $\chi^{2}$ can
be approximated by a quadratic expansion in $\{a_{i}\}$ around
$\left\{ a_{i}^{0}\right\}$.  Once the Hessian is known, one can
estimate not only the uncertainties of $\{a_{i}\}$, but also the
uncertainty in the theoretical prediction for any physical quantity
$X$, provided the dependence of $X$ on $\{a_{i}\}$ can be
approximated by a linear expansion around $\left\{a_{i}^{0}\right\}$,
and is thus characterized by its gradient at
$\left\{a_{i}^{0}\right\}$ (cf.\ Sec.~\ref{sec:ErrorMatrix}).

The first problem we address is a technical one that is important in
practice:
If the uncertainties are very disparate for different directions in the
$n$-dimensional parameter space $\{a_{i}\}$,
{\it i.e.,} if the eigenvalues of $H_{ij}$ span many orders of
magnitude, how can one calculate the matrix $H_{ij}$ with sufficient
accuracy that reliable predictions are obtained for all directions?
To solve this problem, we have developed an iterative
procedure that adapts the step sizes used in the numerical calculation
of the Hessian to the uncertainties in each eigenvector direction.
We demonstrate the effectiveness of this procedure in our specific
application, where the standard tool fails to yield reliable results.

The second problem we address concerns the reliability of estimating
the uncertainty $\Delta{X}$ in the prediction for some
physical variable $X$ that is a function of the $\{a_{i}\}$:  How can
one estimate $\Delta{X}$ in a way that takes into account the
variation of $\chi^{2}$ over the entire parameter space $\{a_{i}\}$,
without assuming the quadratic approximation to $\chi^{2}$ and the
linear approximation to $X$ that are a part of the error matrix approach?
We solve this problem by using
Lagrange's method of the undetermined multiplier to make
constrained fits that derive the dependence of $\chi^2$ on $X$.
Because this method is more robust, it can be used by itself or to
check the reliability of the Hessian method.

Section 2 summarizes the error matrix formalism
and establishes our notation.
Section 3 describes the iterative method for calculating the Hessian,
and demonstrates its superiority in a concrete example.
Section 4 introduces the Lagrange multiplier method and compares its
results with the Hessian approach to the same application.
Section 5 concludes.
%

\section{Error Matrix and Hessian}

\label{sec:ErrorMatrix}

First we review the well-known connection between the error matrix
and the Hessian matrix of second derivatives.
We emphasize the eigenvector representations of those
matrices, which are used extensively later in the paper.

The basic assumption of the error matrix approach is that $\chi^{2}$
can be approximated by a quadratic expansion in the fit parameters
$\{a_{i}\}$ near the global minimum.  This assumption will be true
if the variation of the theory values $T_I$ with $\{a_i\}$ is
approximately linear near the minimum.
Defining $y_{i} = a_{i} - a_{i}^{0}$ as the displacement of parameter
$a_{i}$ from its value $a^{0}_{i}$ at the minimum, we have
\begin{eqnarray}
\chi^{2} &=&\chi_{0}^{\,2}\,+\, \sum_{i,j}H_{ij}\,y_{i}\,y_{j}\,,
\label{eq:taylor}
\\
H_{ij} &=& \frac{1}{2} \left(
\frac{{\partial }^{2}\chi^{2}}{\partial y_{i}\,\partial y_{j}}
\right)_0 \,,
\label{eq:hessian}
\end{eqnarray}
where the derivatives are evaluated at the minimum point {$y_i = 0$}
and $H_{ij}$ are the elements of the
{\it Hessian matrix}.\nolinebreak\footnote{We include a factor $1/2$ in
the definition of $H$, as is the custom in high energy physics.}
There are no linear terms in $y_{i}$ in (\ref{eq:taylor}), because
the first derivatives of $\chi^{2}$ are zero at the minimum.

Being a symmetric matrix, $H_{ij}$ has a complete set of
$n$ orthonormal eigenvectors $V_i^{(k)} \equiv v_{ik}$
with eigenvalues $\epsilon_{k}\,$:
\begin{eqnarray}
\sum_{j}H_{ij}\,v_{jk} &=&\epsilon_{k}\,v_{ik}
\label{eq:Hevec}
\\
\sum_{i}v_{ij}\,v_{ik} &=&\delta_{jk}\,.
\label{eq:orthonormal}
\end{eqnarray}
These eigenvectors provide a natural basis to express arbitrary
variations around the minimum: we replace $\{y_{i}\}$
by a new set of parameters $\{z_{i}\}$ defined by
\begin{equation}
y_{i}=\sum_{j}v_{ij}\,
{\textstyle\sqrt{\frac{1}{\epsilon_{j}}}}\,z_{j} \, .
\label{eq:vij}
\end{equation}
These parameters have the simple property that
\begin{equation}
\Delta \chi^{2}=\chi^{2}-\chi_{0}^{\,2}\,=\,\sum_{i}z_{i}^{\,2}\,.
\label{eq:vij2}
\end{equation}
In other words, {\it the surfaces of constant $\chi^{2}$ are
spheres in $\{z_{i}\}$ space,} with $\Delta \chi^{2}$ {\it the
squared distance from the minimum.}

The orthonormality of $v_{ij}$ can be used to invert the
transformation (\ref{eq:vij}):
\begin{eqnarray}
z_{i}=\sqrt{\epsilon_{i}}\,
\sum_{j}y_{j}\,v_{ji}\,.
\label{eq:zsubi}
\end{eqnarray}
The Hessian and its inverse, which is the error
matrix, are easily expressed
in terms of the eigenvalues and eigenvector components:
\begin{eqnarray}
H_{ij} &=&\sum_{k}\epsilon_{k}\,v_{ik}\,v_{jk} \\
(H^{-1})_{ij} &=&\sum_{k}\frac{1}{\epsilon_{k}}\,v_{ik}\,v_{jk}\,.
\label{eq:Hinv}
\end{eqnarray}

Now consider any physical quantity $X$ that can be calculated
according
to the theory as a function of the parameters $\{a_{i}\}$.  The
best estimate of $X$ is the value at the minimum $X_{0}=X(a_i^0)$.
In the neighborhood of the minimum, assuming the first term of the
Taylor-series expansion of $X$ gives an adequate approximation, the
deviation of $X$ from its best estimate is given by
\begin{equation}
\Delta X = X-X_{0}\,\cong \,\sum_{i}
\frac{\partial X}{\partial y_{i}}\,y_{i}
=\,\sum_{i}X_{i}\,z_{i}
\label{eq:linear}
\end{equation}
where
\begin{equation}
X_{i} \equiv \frac{\partial X}{\partial z_{i}}
\end{equation}
are the components of the $z$-gradient evaluated at
the global minimum, {\it i.e.,} at the origin in $z$-space.

Since $\chi^{2}$ increases uniformly
in all directions in $z$-space,
the gradient vector $X_{i}$ gives the direction in which the
physical observable $X$ varies fastest with increasing $\chi^{2}$.
The maximum deviation in $X$ for a given increase in $\chi^{2}$ is
therefore obtained by the dot product of the gradient vector
$X_{i}$ and a displacement vector $Z_{i}$
in the same direction with length
$\sqrt{\Delta \chi^{2}}$,
{\it i.e.,}\ $Z_{i}=X_{i}\sqrt{\Delta \chi^{2}/\Sigma_{j}X_{j}^{2}}$.
For the square of the deviation, we therefore obtain the simpler formula
\begin{equation}
(\Delta X)^{2}=(X\cdot Z)^{2}=\Delta \chi^{2}\,\sum_{i}X_{i}^{\,2}\,.
\label{eq:DeltaFsq1}
\end{equation}

The traditional formula for the error estimate $(\Delta X)^{2}$ in
terms of the original coordinates $\{y_{i}\}$ can be derived by
substituting $X_{i}=\frac{\partial X}{\partial z_{i}}
={\displaystyle\sum_{j}}\frac{\partial X}{\partial y_{j}}\,
\frac{\partial y_{j}}{\partial z_{i}}$
in (\ref{eq:DeltaFsq1}) and using
(\ref{eq:vij}) and (\ref{eq:Hinv}).
The result is
\begin{equation}
(\Delta X)^{2} \, = \, \Delta \chi^{2}\,\sum_{i,j}
\frac{\partial X}{\partial y_{i}} \, (H^{-1})_{ij}\,
\frac{\partial X}{\partial y_{j}}\;.
\label{eq:DeltaFsq2}
\end{equation}
This standard result can of course also be derived directly by
minimizing $\chi^2$ in (\ref{eq:taylor}) with respect to
$\{a_{i}\}$, subject to a constraint on $X$.

Equations (\ref{eq:DeltaFsq1}) and (\ref{eq:DeltaFsq2})
are equivalent if the assumptions of a linear
approximation for $X$ and a quadratic approximation for $\chi^{2}$ are
exact.  But in practice, the numerical accuracy of the two
can differ considerably if these conditions are not well
met over the relevant region of parameter space.
To calculate the error estimate $\Delta X$, we prefer to use
Eq.\ (\ref{eq:DeltaFsq1}) using derivatives $X_i$ calculated by
finite differences of $X$ at the points
$z_i = \pm \frac{1}{2} \sqrt{\Delta \chi^2}$ (with $z_j = 0$ for $j \ne i$).
This is generally more accurate, because it estimates the necessary
derivatives using an appropriate step size, and thus reduces the effect
of higher order terms and numerical noise.

In a complex problem such as a global analysis, the region of
applicability of the approximations is generally unknown beforehand.
A situation of particular concern is when the various eigenvalues
$\{\epsilon_{i}\}$ have very different orders of
magnitude---signaling that the function
$\chi^{2}$ varies slowly in some directions of ${a_i}$ space, and
rapidly in others.  The iterative method described in the next section
is designed to deal effectively with this situation.
%

\section{Iterative Procedure}

\label{sec:IterativeProcedure}

In practical applications, the Hessian matrix $H_{ij}$
is calculated using finite differences to estimate the
second derivatives in (\ref{eq:hessian}).
A balance must be maintained in choosing the step sizes for this,
since higher-order terms will contribute if the intervals are too large,
while numerical noise will dominate if the intervals are too small.
This noise problem may arise more often than is generally realized,
since the theory values $\{T_{I}\}$
that enter the $\chi^{2}$ calculation may not be the ideally
smooth functions of the fit parameters that one would associate
with analytic formulas.
For in complex theoretical models, the $\{T_{I}\}$ may be
computed from multiple integrals that have small
discontinuities as functions of $\{a_{i}\}$
induced by adaptive integration methods.
These numerical errors forbid the use of a very small step size
in the finite difference calculations of derivatives.
Furthermore, as noted above, the eigenvalues of $H_{ij}$
may span a wide range, so excellent accuracy is needed especially
to get the smaller ones right.

\subsubsection*{The Procedure}

We want to evaluate $H_{ij}$ by sampling the values
of $\chi^{2}$ in a region of parameter space where
Eq.\ (\ref{eq:taylor}) is a good approximation.
In principle, the parameters $\{z_{i}\}$
are the natural choice for exploring this space;
but of course they are not known in advance.
We therefore adopt the following iterative procedure:
\begin{enumerate}
\item  Define a new set of coordinates $\{\xi_i\}$ by
\begin{eqnarray}
y_i = \sum_j u_{ij} \, t_j \, \xi_j
\end{eqnarray}
where $u_{ij}$ is an orthogonal matrix and $\{t_i\}$ are scale
factors.  In the first iteration, these are chosen as
$u_{ij}=\delta_{ij}$ and $t_{i}=1$, so that $\xi_{i}=y_{i}$.
This makes the first round of iteration similar to the
usual procedure of taking derivatives with respect to $a_i\,$.
The iterative method is designed such that with successive iterations,
$u_{ij}$, $t_{i}$, and $\xi_{i}$ converge to
$v_{ij}$, $\sqrt{1/\epsilon_{i}}$, and $z_{i}\,$ respectively.

\item  Calculate the effective second derivative matrix $\Phi_{ij}$
defined by
\begin{eqnarray}
\chi^{2} &=&\chi_{0}^{\,2} \, + \,
\sum_{ij}\Phi_{ij}\,\xi_{i}\,\xi_{j} \\
\Phi_{ij} &=&\frac{1}{2} \, \frac{{\partial }^{2}\chi^{2}}
{\partial \xi_{i}\,\partial \xi_{j}}
\end{eqnarray}
using finite differences of the $\xi_{i}$.
The step size in $\xi_{i}$ is chosen to make the increase in
$\chi^{2}$ due to the diagonal element
$\Phi_{ii} \xi_i^{2}$
equal to a certain value $\delta\chi^{2}\,$.
The choice of $\delta\chi^{2}$ is determined by the particular
physics application at hand.  Naively, one might expect
$\delta\chi^{2} \simeq 1$ to be the right choice.
That would indeed be appropriate for a $\chi^{2}$
function obeying ideal statistical requirements.
But when the input to $\chi_{\rm global}^{2}$ is imperfect,
a reasonable choice of $\delta \chi^{2}$ must be based on a
physics judgement of the appropriate range of that
particular $\chi^2$ function.
We therefore leave the choice of $\delta \chi^2$ open in this
general discussion.\nolinebreak\footnote{Cf.\ discussion in the
following subsection on a sample problem.}
In any case, if the final results are
to be trustworthy, they must not be sensitive to that choice.

We calculate each off-diagonal second derivative by evaluating
$\chi^{2}$ at the four corners of the rectangle
$(+\delta_{i},+\delta_{j})$,
$(-\delta_{i},-\delta_{j})$,
$(+\delta_{i},-\delta_{j})$,
$(-\delta_{i},+\delta_{j})$, where $\delta_{i}$ is the step size.
This is a modification of the technique
used in {\footnotesize MINUIT} \cite{minuit}.
For the sake of efficiency, the {\footnotesize MINUIT} subroutine
{\footnotesize HESSE} estimates off-diagonal elements
using only one of those corners, together with values at
$(\delta_{i},0)$ and
$(0,\delta_{j})$ that are already known from
calculating the diagonal elements of the Hessian.
Our method is slower by a factor of 4, but is more accurate
because it fully or partly cancels some of the contributions from
higher derivatives.
The first derivatives $\partial \chi^{2}/\partial \xi_{i}$
are also calculated at this stage of the iteration and used to
refine the estimate of the location of the minimum.

\item  Compute the Hessian according to $\Phi_{ij}$,
\begin{equation}
H_{ij}=\sum_{m,n}\frac{\Phi_{mn}\,u_{im}\,u_{jn}}{t_{m}\,t_{n}}\,.
\end{equation}

\item  Find the normalized eigenvectors of the Hessian, as
defined by Eqs.\ (\ref{eq:Hevec}) and (\ref{eq:orthonormal}).

\item  Replace $u_{ij}$ by $v_{ij}$, $t_{j}$ by
$\sqrt{1/\epsilon_{j}}\,$, and go back to step 1.
The steps are repeated typically 10--20 times, until the
changes become small and $\Phi_{ij}$ converges to $\delta_{ij}$.
\end{enumerate}

This iterative procedure improves the estimate of the Hessian matrix,
and hence of the error matrix, because in the later iterations it
calculates the Hessian based on points that sample the region where
$\Delta\chi^{2}$ has the magnitude of physical interest.

\subsubsection*{Results from a Sample Application}

As an example, we apply the iterative procedure to the application
that motivated this study---the global analysis of
PDFs \cite{runII}---and compare the results with those obtained from
{\footnotesize MINUIT}.
The experimental input for this problem consists of
$N \! = \! 1295$ data points from 15 different experimental data
sets involving four distinct physical processes.
All the potential complexities mentioned earlier are present in this
system.  The theory is the quark parton model, based on next-to-leading
order perturbative Quantum Chromodynamics (QCD).  The model contains
$n=16$
parameters $a_{i}$ that characterize the quark and gluon distributions
in the proton at some low momentum scale $Q_{0}$.
From a calculational point of view, the theoretical model consists
of the numerical integration of an integro-differential equation and
multiple convolution integrals that are evaluated mostly by adaptive
algorithms.
The fitting function $\chi_{\rm global}^{2}$ in this case combines
the published statistical and systematic errors of the data points
in quadrature.
The only correlated errors incorporated are the published overall
normalization uncertainties of the individual experiments.
The fitting program is the same as that used to generate the
CTEQ parton distributions \cite{CTEQn, cteq5}.
The global $\chi^{2}$ minimum for this system
defines the CTEQ5M1 set of PDFs,
for which $\chi_{0}^{2}\approx 1200$ \cite{cteq5}.
We find that the eigenvalues $\{\epsilon_{i}\}$ of the Hessian
for this system range over 5--6 orders of magnitude (distributed
approximately exponentially).

The value of $\Delta\chi^{2}$ that corresponds to a given confidence
level is well defined for an ideal experiment:  {\it e.g.,}
$\Delta\chi^{2} < 1$ defines the $68\%$ confidence region.
But in a real-world global analysis, the experimental and theoretical
values in Eq.~(\ref{eq:ChisqDef}) include systematic errors, and
the uncertainties $\sigma_{I}$ include subjective estimates of those
errors, so the relation between $\Delta\chi^{2}$ and confidence level
requires further analysis.
From independent detailed studies of the
uncertainties \cite{LMpaper,Hessepaper},
we estimate that an appropriate choice of $\delta\chi^{2}$ for the
iterative calculation is around $10$ in our application, and
only the region $\Delta\chi^{2} > 100$ can be ruled out for the
final fits.

The error matrix approach relies on a quadratic approximation
to $\chi^{2}$ in the neighborhood of the minimum.
To test whether that approximation is valid, we plot $\chi^{2}$
as a function of distance along a particular direction
in $\{a_{i}\}$ space, as shown in Fig.\ \ref{fig:one}.
The direction chosen is a typical one---specifically it is the
direction of the eigenvector with median eigenvalue.
The dotted curve in Fig.\ \ref{fig:one} is the exact
$\chi^{2}$ and the solid curve is the quadratic approximation
(\ref{eq:taylor}).
The approximation is seen to provide a rather good
description of the function.
Even at points where $\chi^{2}$ has increased by $50$,
the quadratic approximation reproduces the increase to
$20$\% accuracy.

\figone

\figtwo

To correctly measure the curvature of the quadratic approximation,
it is important to fit points that are displaced by an appropriate
distance.  This can be seen from Fig.\ \ref{fig:two}, which shows
the difference between the two curves in Fig.\ \ref{fig:one} in
the central region.
The difference displays a small cubic contribution to $\chi^{2}$.
It also reveals contributions that vary erratically
with a magnitude on the order of $0.03\,$.
These fluctuations come from the noise associated with
switching of intervals in the adaptive integration routines.
Because the fluctuations are small, they do not affect our results
in principle.
But they do require care in estimating the derivatives.
In particular, they would make finite-difference estimates
based on small intervals extremely unreliable.
The iterative method avoids this problem by choosing
a suitable scale for each eigenvector direction when evaluating
the Hessian.

Figures \ref{fig:one} and \ref{fig:two} show the behavior of
$\chi^{2}$ along a single typical direction in the 16
dimensional parameter space.
Fig.\ \ref{fig:three} shows a complementary test of the iterative
method for all possible directions.
We have chosen 1000 directions at random in $\{z_{i}\}$ space.
We displace the parameters away from the minimum
in each of these directions by a distance that makes
$\Delta\chi^{2}=5$.
We then compute the value of $\Delta\chi^{2}$ predicted
by the quadratic approximation (\ref{eq:taylor}), using the Hessian
calculated by the iterative method and, for comparison, by the
routine {\footnotesize HESSE} within the {\footnotesize MINUIT}
package.
The results are displayed in Fig.\ \ref{fig:three}
as histograms, with $\Delta\chi^{2}$
on the horizontal axis and the number of counts on the vertical axis.
If $\chi^{2}$ were quadratic in $\{a_{i}\}$, then
a perfect computational method would yield a delta function
at $\Delta\chi^{2}=5$.
Fig.\ \ref{fig:three} shows that:

\begin{Simlis}{1em}
\item  For the {\it solid histogram}---the result of the iterative
procedure---the quadratic approximation is close to the
exact result in all directions, and hence Eq.\ (\ref{eq:taylor})
is a pretty good representation of $\chi^{2}$.
Quantitatively, the middle $68$\% of the distribution is contained
in the region $5.4 \pm 0.6\,$.

\item  For the {\it dotted histogram}---based on the general purpose
program {\footnotesize MINUIT}---the distribution is also spread around
the expected value of $5$, but it is very broadly distributed.
This estimate of the Hessian is therefore unsatisfactory,
because we might be interested in a quantity whose gradient direction
is one for which the Hessian computed by {\footnotesize MINUIT}
is widely off the mark.
A major source of this problem is the numerical noise visible
in Fig.\ \ref{fig:two}: {\footnotesize MINUIT} uses a small step
size to calculate the derivatives, and gets misled by the small-scale
discontinuities in $\chi^{2}$.
For some directions, $\Delta\chi^{2}$ even becomes negative because
the errors in one or more of the small eigenvalues are big enough to
allow their calculated values to become negative.
(Within {\footnotesize MINUIT}, this circumstance elicits a
warning message, and a constant is added to all the eigenvalues, which
in the context of Fig.\ \ref{fig:three} corresponds to shifting
the dotted distribution to the right.)
\end{Simlis}

\figthree

\figfour

Figure \ref{fig:four} shows the results of a similar study,
in which the 1000 random directions are chosen only from the
subspace of $\{z_{i}\}$ that is spanned by the 10 directions with
the largest eigenvalues $\epsilon_{i}$.
The larger eigenvalues correspond to directions in which $\chi^{2}$
rises most rapidly, or in other words, directions in which the
parameters are more strongly constrained by data.
Because the distance moved away from the minimum in $\{a_{i}\}$
space is smaller in this case, the quadratic approximation is
generally better, so it is not surprising that the
histograms are more sharply peaked than in Fig.\ \ref{fig:three}.
But the advantage of the iterative method remains apparent.

\subsubsection*{Comment}
Information from the iteratively-improved Hessian provides a useful
tool for refining the choice of functional forms used to
parametrize a continuous degree of freedom in the theoretical model.
Explicitly, the relevant formulas are as follows.

The length squared of the displacement vector in the space of fit
parameters is
\begin{eqnarray}
\sum_i (a_i - a_i^0)^{2} = \sum_i y_i^{\, 2} =
\sum_i \frac{z_i^{\, 2}}{\epsilon_i}
\label{eq:lengthsq}
\end{eqnarray}
while
$\Delta \chi^{2}=\sum_{i} z_{i}^{\,2}$ by
(\ref{eq:vij2}).  Hence the directions in which the parameters
are well determined (the steep directions) correspond to eigenvectors
of the Hessian
with large eigenvalues, while the shallow directions in which
they are weakly determined correspond to small eigenvalues.

The extreme values for any particular $a_i$ are
\begin{eqnarray}
a_{i} = a_{i}^{0} \, \pm \, \Delta a_{i}  \label{eq:extremeai}
\end{eqnarray}
where
\begin{eqnarray}
(\Delta a_{i})^{2} = \Delta \chi^2 \, \sum_j
\frac{v_{ij}^{\, 2}}{\epsilon_{j}} \,.
\label{eq:deltaai}
\end{eqnarray}
Equation (\ref{eq:deltaai}) can be used to see if each parameter is
appropriately well constrained.  Furthermore, the individual terms
in the sum show the contributions to $\Delta a_i$ from the various
eigenvectors, so if a parametrization leads to a poorly defined
minimum because it allows too much freedom---which is indicated
by a failure of the iteration to converge for the smallest eigenvalues
of the Hessian---it is easy to see which of the parameters are most
responsible for the too-shallow directions.
%

\section{Lagrange Multiplier Method}

The Hessian, {\it via} its inverse which is the error
matrix,
provides a general way to propagate the uncertainties
of experimental and theoretical input to the fit parameters
$\{a_{i}\}$, and thence on to a measurable quantity
$X(\{a_{i}\})$ by Eqs.\ (\ref{eq:DeltaFsq1})
or (\ref{eq:DeltaFsq2}).
But these equations are based on assuming that $\chi^{2}$ and $X$
can be treated
as quadratic and linear functions of $\{a_{i}\}$ respectively.
In this section we describe a different approach, based
on the mathematical method of the {\it Lagrange undetermined
multiplier}, which avoids those assumptions.

\subsubsection*{The Procedure}
Let $X_{0}$ be the value of $X$ at the $\chi^{2}$ minimum, which
is the best estimate of $X$.
For a fixed value of $\lambda$, called the Lagrange multiplier,
one performs a new minimization with respect to the fit parameters
$\{a_{i}\}$, this time on the quantity
\begin{equation}
F=\chi^{2}\,+\,\lambda (X-X_{0})\,,
\label{eq:DefineF}
\end{equation}
to obtain a pair of values $(\chi^{2}(\lambda),X(\lambda))$.
(The constant term $-\lambda X_{0}$ here is not necessary,
because it does not affect the minimization; but it makes the
minimum value of $F$ easier to interpret.)
At this new minimum,
$\chi^{2}(\lambda)$ is the lowest possible $\chi^{2}$
for the corresponding value $X(\lambda)$ of the physical
variable $X$.
Thus one achieves a {\it constrained fit}
in which $\chi^{2}$ is minimized for a particular
value of $X$.

By repeating the minimization for many values of $\lambda$, one
maps out the parametrically-defined curve
$(\chi^{2}(\lambda),X(\lambda))$.
Since $\lambda$ is just the parameter for this curve,
its value is of no particular physical significance.
The relevant range for $\lambda$ can be found by trial-and-error;
or it can be estimated using the Hessian approximation, which
predicts that
$\lambda \approx -2\,\Delta \chi^{2}/\Delta X$.
In that approximation, $F$ goes down by the same amount
that $\chi^{2}$ goes up.

One way to understand the Lagrange Multiplier method is to
imagine that the quantity $X$ is simply one of the fitting
parameters, say $a_{1}$.  The variation of $\chi^{2}$ with $a_{1}$
could be mapped out by minimizing $\chi^{2}$ with respect to
$\{a_{2},\dots,a_{n}\}$ for a sequence of values of $a_{1}$.
(That operation is indeed so useful that {\footnotesize MINUIT}
provides a procedure {\footnotesize MINOS} to carry it out.)
In the more general case that $X$ is a function of all $\{a_{i}\}$,
one wants to similarly minimize $\chi^{2}$ for fixed values of $X$.
That is exactly what the Lagrange Multiplier method does,
since including the undetermined multiplier term in
(\ref{eq:DefineF}) renders the $\{a_{i}\}$
independent in spite of the constraint on $X$.

A more phenomenological way to understand the Lagrange
Multiplier method is to imagine that $X$ has just been measured,
with result $X_{\mathrm{new}}\pm \sigma_{\mathrm{new}}$.
To decide whether this hypothetical new measurement is consistent
with the old body of data, one would add a term
$[(X_{\mathrm{new}}-X)/\sigma_{\mathrm{new}}]^{2}$
to $\chi_{\rm global}^{2}$ of Eq.\ (\ref{eq:ChisqDef}) and redo
the minimization.  The added contribution to $\chi^{2}$ consists
of a constant, a linear term in $X$, and a quadratic term in $X$.
This is equivalent to Eq.\ (\ref{eq:DefineF}), because a
constraint on $X^{2}$ is equivalent to a constraint on $X$ itself.

The essential feature of the Lagrange Multiplier method is that,
for a given $\Delta\chi^{2}$, it finds the
largest range of $X$ allowed by the global data set
and the theoretical model, independent of any approximations.
The full parameter space $\{a_{i}\}$ is explored
in the minimization procedure, not just the immediate neighborhood
of the original $\chi^{2}$ minimum as in the Hessian method,
and no approximations based on a small deviation from the original
minimum are needed.

The only drawback to the Lagrange Multiplier method is that it can
be slow computationally, since it requires a separate series of
minimizations for each observable $X$ that is of interest.

\subsubsection*{Example}
We now look at an example of the Lagrange Multiplier method from
our application, the uncertainty of parton distribution functions.
For the physical quantity $X$, we consider the cross section
$\sigma_{W}$ for $W^{\pm}$ production in $p\overline{p}$ collisions
at the energy $\sqrt{s}=1.8$\,TeV of the Tevatron collider at
Fermilab.
We want to estimate $\sigma_{W}$
and the uncertainty on that estimate,
based on the global analysis of parton distributions.

The points in Fig.\ \ref{fig:five} show $\chi_{\rm global}^{2}$
as a function of $\sigma_{W}B$ in nanobarns, where
$B=0.106$ is the branching ratio assumed for $W\rightarrow e\,\nu$.
These points are obtained by the Lagrange Multiplier method using
$\lambda=0$, $\pm 1000$, $\pm 2000$, $\pm 3000$, $\pm 4000$.
They are thus discrete cases of $\chi^{2}_{\rm global}$
versus $\sigma_{W}$, without approximations.

\figfive

The smooth curve in Fig.\ \ref{fig:five} is the parabola
given in Eq.\ (\ref{eq:DeltaFsq2}), using
the Hessian computed by the iterative method and treating
$\sigma_{W}$ in the linear approximation.
The comparison between this curve and the discrete points from the
Lagrange Multiplier calculation tests the quality of the
quadratic and linear approximations and the reliability of
the iterative calculation of $H_{ij}$.
For this application, we conclude that the improved Hessian method
works very well, since the difference between the points and the
curve is small, and indicates only a small cubic
contribution.  If the two results
did not agree, the correct result would be the one given by the
Lagrange Multiplier method.

To estimate $\Delta X$, the uncertainty of $X$ consistent with the
global analysis of existing data, one needs to specify what range
of $\Delta\chi_{\rm global}^{2}$ is allowed.
As discussed earlier, the acceptable limit of
$\chi^{2}_{\rm global}$ depends on the nature of the original
definition of this fitting function, in the context of the
specific system under study.
For the case of $\sigma_{W}$, Ref.\ \cite{Hessepaper} estimates
$\Delta\chi_{\rm global}^{2}\approx 100$,
which translates into $\Delta\sigma_{W}/\sigma_{W}\approx \pm 3$\%
according to Fig.\ \ref{fig:five}.
%

\section{Conclusion}
\label{sec:Conclusion}

We have addressed some computational problems that arise in a
global phenomenological analysis, in which a complex theory with
many parameters confronts a large number of data points from
diverse experiments.

The traditional error-matrix analysis is based on a quadratic
approximation to the function $\chi^{2}$ that measures the quality
of the fit, in the neighborhood of the minimum that defines the
best fit.
The iterative method proposed in Sec.\ 3 improves the
calculation of the Hessian matrix which expresses that quadratic
approximation, for a
complex system in which general-purpose programs may fall short.
The inverse of this improved version of the Hessian matrix is an
improved version of the error matrix.  It can be used to estimate
the uncertainty of predictions using standard error matrix formulas.

Our iterative procedure for calculating the Hessian is implemented
as an extension to the widely-used CERN fortran library routine
{\footnotesize MINUIT} \cite{minuit}.
The code is available from
http://www.pa.msu.edu/${\scriptstyle \sim}$pumplin/iterate/,
or it can be requested by e-mail from
\newline
pumplin@pa.msu.edu.
Included with this code is a test example which demonstrates that
the iterative method is superior to standard {\footnotesize MINUIT}
{\em even for a $\chi^2$ function that has no numerical noise of
the type encountered in Fig.\ \ref{fig:two}}.

The Lagrange Multiplier method proposed in Sec.\ 4 calculates the
uncertainty on a given physical observable directly, without
going through the error matrix.  It thus avoids the assumption that
the theoretical quantities can be approximated by linear functions
of the search parameters, which is intrinsic to the Hessian approach.

For simplicity, we have discussed only the problem of obtaining
error estimates on a single quantity $X$.  It is straightforward
to generalize our methods to find the region allowed simultaneously
for two or more variables by a given $\Delta\chi^{2}$.
For example, in the case of two variables $X^{(1)}$ and $X^{(2)}$,
the allowed region according to the Hessian method is the interior
of an ellipse.  The Lagrange multiplier method can be generalized for
this case by adding {\it two} terms,
$\lambda_1 X^{(1)} \, + \, \lambda_2 X^{(2)}$, to $\chi^2$.

Although the Lagrange Multiplier procedure is conceptually simple
and straightforward to implement, it is slow computationally because
it requires many full minimizations to map out $\chi ^{2}$ as a
function of $X$, and this must be done separately for each quantity
$X$ whose error limits are of interest.
In contrast, once the Hessian has been determined from the
global analysis, it can be applied to any physical observable.
One needs only to compute the gradient
${\partial X}/{\partial a_{i}}$ of the observable $X$ and substitute
into Eq.\ (\ref{eq:DeltaFsq2}); or better, to compute the gradient
$X_i = {\partial X}/{\partial z_{i}}$ and substitute into
Eq.\ (\ref{eq:DeltaFsq1}).
For computational efficiency, the iteratively calculated Hessian
is therefore the method of choice, provided its linear
approximations
are sufficiently accurate.  Whether or not that is the case can be
determined by comparing Hessian and Lagrange Multiplier results.
We use both methods in a detailed study of the uncertainties
in the CTEQ5 parton distribution
functions \cite{runII,LMpaper,Hessepaper} that is based on the work
presented here.

\section*{Acknowledgments}

We thank R.~Brock, D.~Casey, J.~Huston, and F.~Olness for
discussions on the uncertainties of parton distribution
functions.  We thank M.~Botje for comments on an earlier
version of the manuscript.
This work was supported in part by NSF grant PHY-9802564.

\end{document}